\documentclass[aps,floats,twocolumn,prl,nofootinbib]{revtex4-2}
\usepackage{amsfonts,amsmath,amssymb,ascmac,bm,amsthm}
\usepackage{fnpct} 
\usepackage{comment}
\usepackage{caption}
\usepackage{ifpdf}
\usepackage{graphicx}
\usepackage{slashed}
\usepackage{color}
\usepackage[mathscr]{eucal}
\usepackage[utf8]{inputenc}
\usepackage{physics}
\usepackage{cancel}
\usepackage{float}
\usepackage{soul}
\usepackage{booktabs}
\usepackage{simpler-wick}
\usepackage{hyperref}
\usepackage{tensor}
\usepackage{upgreek}
\usepackage{caption}
\usepackage{subfigure}
\usepackage{subcaption}
\usepackage{tikz}
\usepackage{appendix}

\newcommand\mi{\mathrm{i}}
\newcommand\me{\mathrm{e}}

\newcommand\pp{\uppi}

\newcommand{\dif}{\mathrm{d}}

\begin{document}

\title{Waveform stability of black hole ringdown with stochastic horizon structure}

\author{Han-Wen Hu$^{1,2}$}
\email{huhanwen@itp.ac.cn}
\author{Cheng-Jun Fang$^{1,2}$}
\email{fangchengjun@itp.ac.cn}
\author{Zong-Kuan Guo$^{1,2,3}$}
\email{guozk@itp.ac.cn}
	\affiliation{$^1$Institute of Theoretical Physics, Chinese Academy of Sciences, P.O. Box 2735, Beijing 100190, China}
	\affiliation{$^2$School of Physical Sciences, University of Chinese Academy of Sciences, No.19A Yuquan Road, Beijing 100049, China}
   \affiliation{$^3$School of Fundamental Physics and Mathematical Sciences, Hangzhou Institute for Advanced Study, University of Chinese Academy of Sciences, Hangzhou 310024, China}

\begin{abstract}
We examine the robustness of black hole ringdown to stochastic horizon-scale structure within an effective field framework in a proof-of-principle Schwarzschild setup.
Consistent with the understanding that the spectral instability of quasinormal modes does not necessarily imply observational breakdown, our results demonstrate that the macroscopic gravitational waveform remains robust. 
We identify the phase averaging mechanism as the physical origin of this stability, demonstrating that the spatial integration of the wave equation efficiently attenuates ultraviolet geometric details below the resolution limit of the probing wavelength. 
Building on the scaling law $\mathcal{M} \propto \epsilon^2$ and the characteristic mismatch profile with respect to $L_c$, we propose a geometric selection rule for observability: a detectable signal imposes a strict dual constraint requiring both macroscopic spatial coherence ($L_c \sim M$) and classical-level intensity ($\epsilon \gtrsim 10^{-4}$). 
This criterion quantitatively rules out the observability of incoherent, high-entropy quantum foam in the present static Schwarzschild model, suggesting that any significant ringdown deviation would instead serve as evidence for macroscopically coherent horizon structures.
\end{abstract}

\maketitle

\section{Introduction}

The direct detection of gravitational waves (GWs) from binary black hole (BH) coalescences by the LIGO-Virgo-KAGRA collaboration has inaugurated a revolutionary era of precision gravity \cite{LIGOScientific:2016aoc,LIGOScientific:2016vbw,Isi:2019aib,LIGOScientific:2025obp,Berti:2025hly}. 
At the heart of this new frontier lies BH spectroscopy, a method that utilizes the quasinormal mode (QNM) spectrum of the ringdown signal to test the no-hair theorem of general relativity (GR) \cite{Dreyer:2003bv,CalderonBustillo:2020rmh,Meidam:2014jpa,Bhagwat:2019dtm,LIGOScientific:2025obp}. 
In the classical GR  framework, a perturbed Kerr BH relaxes into a stable state by radiating GWs characterized by complex frequencies $\omega$, which are uniquely determined by the remnant's mass and spin \cite{Bhagwat:2019dtm,Volkel:2025jdx,Meidam:2014jpa,Isi:2019aib}. 
While current observations of the dominant mode and potential overtones remain consistent with GR predictions \cite{Bhagwat:2019dtm,Berti:2025hly}, the quest for higher-order modes and potential deviations continues to drive the field toward the upcoming third-generation ground-based detectors and space-based missions like LISA.

However, the theoretical purity of the classical horizon is challenged by the BH information paradox and various quantum gravity proposals \cite{Mathur:2008nj,Saravani:2012is,Susskind:2012rm,Harlow:2013tf,Devin:2014sma}. 
Models such as fuzzballs and firewalls suggest that the classical horizon may be replaced by a structured structure \cite{Mathur:2005zp,Mathur:2008nj,Bena:2022rna,Mayerson:2022yoc,Ikeda:2021uvc,Mathur:2014nja,Oshita:2016pbh}. 
While coherent microstate geometries can break classical symmetries and possess rich multipolar structures \cite{Bianchi:2020bxa}, from the perspective of dynamical scattering, these complex boundaries may manifest as effective potential roughness or reflectivity \cite{Cardoso:2016oxy,Cardoso:2017cqb,Oshita:2018fqu,Abedi:2016hgu,Abedi:2020sgg,Cao:2025qxt}.
Despite intensive searches, conclusive evidence for such deviations, such as GW echoes, remains elusive in current data \cite{Abedi:2020sgg,Lo:2018sep,Tsang:2019zra,Abedi:2022bph}. 
This observational void compels us to understand not just the statics \cite{Conklin:2017lwb,Ren:2021xbe,Hu:2025beh}, but the dynamical stability of ringdown signals against such horizon-scale modifications.

Parallel to these debates, another crisis has emerged, casting doubt on the stability of QNMs \cite{Jaramillo:2020tuu,Boyanov:2023qqf,Shen:2025yiy,Oshita:2025ibu,Cai:2025irl,Destounis:2023nmb,Cao:2025afs}. 
The QNM spectrum of a Schwarzschild BH is known to be exponentially sensitive to infinitesimal perturbations of the potential barrier \cite{Nollert:1996rf,Cheung:2021bol,Hu:2025efp}.
Such ``spectral instability" suggests that the discrete pole distribution in the complex plane could be completely rearranged by both random perturbations and deterministic high-frequency structures of the form $\delta \tilde V_d \propto \cos(2\pp k x)$, potentially invalidating the entire framework of BH spectroscopy \cite{Jaramillo:2021tmt,Jaramillo:2020tuu}. 
This leads to a profound paradox: if the spectrum is mathematically unstable, why do numerical simulations and astronomical observations consistently retrieve the standard Kerr QNMs \cite{Cai:2025irl}? 
While it is recognized that spectral instability does not directly equate to waveform instability \cite{Nollert:1996rf,Xie:2025jbr,Wu:2025sbq}, the mechanism that bridges this gap in the presence of realistic quantum effects remains poorly understood.

In this work, as a proof-of-principle study, we systematically investigate the robustness of BH ringdown to a static stochastic correction of the Schwarzschild barrier using an effective field description. 
By coupling the linearized Einstein equations to a stochastic source \cite{Verdaguer:2000md,Hu:2008rga,Rodriguez:2025xly}, we model the geometric corrections as random potential perturbations defined by the fluctuation intensity $\epsilon$ and the correlation length $L_c$.
Our analysis identifies the phase averaging mechanism as the physical origin of waveform robustness, effectively bridging the gap between mathematical spectral instability and observational stability. 
We demonstrate that the ultraviolet geometric details below the resolution limit of the probing wavelength ($L_c \ll M$) are smoothed out by the wave propagation integral. 
Consequently, we propose a geometric selection rule for observability: a detectable signal imposes a strict dual constraint requiring both macroscopic spatial coherence ($L_c \sim M$) and classical-level intensity. 
This criterion suggests that incoherent, high-entropy quantum foam remains observationally indistinguishable from the classical background, and any significant ringdown deviation would serve as evidence macroscopically coherent horizon structures.

\section{Stochastic description for horizon structure}

To quantitatively describe the backreaction of horizon structure, we adopt a stochastic field approach within linearized gravity. 
The governing equation takes a form structurally identical to the Einstein-Langevin equation
\cite{Hu:2008rga}:
\begin{equation}\label{eq:ELE}
    G_{\mu \nu}[g+h] =\langle T_{\mu \nu}[g+h] \rangle + \xi_{\mu \nu}[g].
\end{equation}
Here, $g\equiv g_{\mu \nu}$ denotes the background metric, $h\equiv h_{\mu \nu}$ represents the induced stochastic metric perturbation, and $\langle T_{\mu \nu} \rangle$ represents the expectation value of the energy-momentum tensor. 
While this formalism was originally developed to describe Planck-scale quantum vacuum noise in semiclassical gravity, we employ it here as an effective field theory description.
In our phenomenological framework, the source term $\xi_{\mu \nu}$ is generalized from a quantum stress-energy fluctuation to an effective stochastic source arising from the statistical variance of the underlying horizon microstates.

By linearizing Eq.\ \eqref{eq:ELE}, we obtain the linear driving equation for the dynamics of the metric fluctuation $h_{\mu \nu}$,
\begin{equation}
    G_{\mu\nu}^{(1)}[h]=\langle T_{\mu \nu}[h] \rangle + \xi_{\mu \nu}[g],
\end{equation}
where the superscript $(1)$ denotes the linear order of Einstein tensor.
For a Schwarzschild BH background, we can neglect the $\langle T_{\mu \nu}[h] \rangle$ term induced by the Hawking radiation. 
Under the Eddington-Finkelstein coordinates $(v, r, \theta, \varphi)$, we introduce a mass function $m(v, r) = M + \eta(v, r)$, where $\eta$ represents the stochastic perturbation of the total mass. 
The metric component $f(r)$ is accordingly rewritten as
\begin{equation}
    f(r) = 1 - \frac{2(M + \eta(v, r))}{r} \equiv f_0(r) + \delta f(v, r).
\end{equation}
According to the dynamical evolution in stochastic gravity, the evolution of the mass fluctuation $\eta$ is driven by the energy flux component $\xi_v^r$
\begin{equation}
    \frac{\partial \eta(v, r)}{\partial v} = 4\pp r^2 \xi_v^r(v, r).
\end{equation}
One can obtain that
\begin{equation}\label{eq:delta-f-vr}
    \delta f(v,r) = -8 \pp r \int^{v}\dif v^{\prime}\  \xi_{v^{\prime}}^r(v^{\prime}, r).
\end{equation}
From the perspective of stochastic analysis, if $\xi_v^r$ is treated as a high-frequency noise process in time, its integral $\eta(v, r)$ essentially follows a Wiener process.
Within the observation timescale of ringdown $\Delta v \sim M$, the variance drift of the Wiener process $\langle \delta f^2 \rangle \propto \Delta v$ is relatively negligible. 
In a given single observation event, Eq.\ \eqref{eq:delta-f-vr} represents a deterministic, non-zero realization of the random field in radial coordinate $r$.  
Consequently, we adopt the quenched disorder approximation, treating $\delta f(v, r)$ as a quasi-static random field $\delta f(r)$.

This stochastic description is theoretically well-motivated even for deterministic quantum geometries. 
In the spirit of random matrix theory applied to black hole scattering \cite{Cotler:2016fpe,Das:2023ulz}, the intricate diffraction effects from a complex, deterministic boundary are effectively indistinguishable from a realization of a random potential. 
Furthermore, within the corpuscular picture of black holes \cite{Dvali:2011aa}, the background geometry emerges from a condensate with a macroscopic occupation number $N \sim (M/l_{\rm P})^2 \gg 1$. 
The collective envelope of these mass fluctuations evolves on a timescale far exceeding the ringdown duration, thereby rigorously validating the quasi-static assumption employed in our model.

\begin{figure}[!htb]
    \centering
    \includegraphics[width=0.9\linewidth]{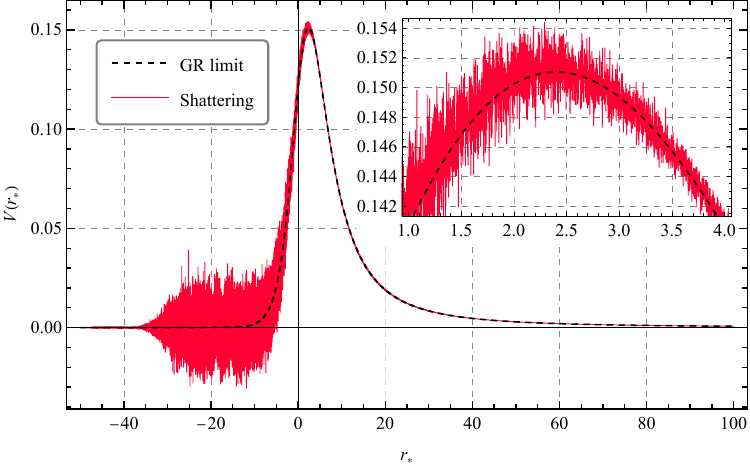}
    \caption{Stochastic potential shattering on a Schwarzschild BH background with $\epsilon = 10^{-3}$ and $L_c = M$. The inset highlights the dense local extrema near the potential peak, where violet fluctuations (red line) break the smooth classical Regge-Wheeler barrier (dashed black line).}
    \label{fig:shattering-potential}
\end{figure}

To describe the spatial distribution of metric fluctuations, we first define a white noise $W(r_*)$ in tortoise coordinates $r_*$ and convolve it with a Gaussian kernel to generate a colored noise field $\zeta(r_*)$
\begin{equation}\label{eq:convolve}
\zeta(r_*) = \mathcal{N} \int_{-\infty}^{+\infty} \dif \tau \ W(\tau) \exp\left( -\frac{(r_*- \tau)^2}{2 L_c^2} \right),
\end{equation}
where $\mathcal{N}$ is a normalization factor.
Considering that quantum gravity effects are typically coupled with local curvature, we assume the intensity of metric perturbations is proportional to the tidal force. 
In a Schwarzschild spacetime, the tidal force is determined by the Kretschmann scalar, whose radial scaling suggests that the perturbation envelope should be proportional to $(2M/r)^3$. 
Thus, the explicit physical form of the metric correction $\delta f(r)$ is defined as
\begin{equation}
\delta f(r) = \epsilon \left( \frac{2M}{r(r_*)} \right)^3 \zeta(r_*),
\end{equation}
where $\epsilon$ and $L_c$ serve as the two fundamental control parameters defining the parameter space of the horizon structure. 
The amplitude $\epsilon$ governs the fluctuation intensity, while the correlation length $L_c$ quantifies the spatial coherence scale: the microscopic limit $L_c \ll M$ corresponds to incoherent quantum foam, whereas $L_c \sim M$ mimics macroscopically coherent geometric deformations
\footnote{
However, the subsequent scattering behavior is controlled primarily by $\epsilon$ and $L_c$, rather than by the detailed shape of the smoothing kernel. 
We have checked that replacing the Gaussian correlator in Eq.\ \eqref{eq:convolve} by other short-range kernels, while matching the same variance and $L_c$, leaves the characteristic mismatch profile at the qualitative level. 
The Gaussian choice is therefore adopted only as a convenient representative parametrization.
}
. 
This allows us to map the $\delta f(r)$ to an effective correction of the Regge-Wheeler potential $\delta V_{\rm eff}(r)$ in the master equation of BH perturbation,
\begin{equation}\label{eq:delta-V}
    \delta V_{\rm eff}(r) = \frac{6 f_0(r)}{r^2} \frac{\dif \eta}{\dif r} - \frac{2 \eta}{r^3} \left[\ell(\ell+1) + 3f_0(r) - \frac{6M}{r} \right].
\end{equation}
The first term in Eq.\ \eqref{eq:delta-V} contains $\dif\eta/\dif r$ is the physical origin of the shattering of the potential barrier at microscopic scales. 
Comparing Eq.\ \eqref{eq:delta-V} with the additional structure of effective potential perturbation in Ref.\ \cite{Jaramillo:2020tuu}, one can obtain that the ratio $\epsilon/L_c$ plays the role of the perturbation wavenumber $k$ in $\delta \tilde{V}_d$, while $\epsilon$ controls the perturbation amplitude or norm.
In contrast, the second term, though containing $\eta$, remains relatively smooth in long-wavelength scattering.
The visual impact of this stochastic correction on the classical Regge-Wheeler barrier is illustrated in Fig.\ \ref{fig:shattering-potential}, which explicitly depicts the potential shattering phenomenon.

Finally, to ensure ingoing wave boundary condition in the numerical evolution and simulate gravitational redshift effects \cite{Gundlach:1993tp,Cardoso:2016oxy}, we introduce a horizon smoothing operator. 
Using a hyperbolic tangent function $\mathcal{F}(r_*)$, we force the stochastic perturbations to vanish smoothly as $r_* \to -\infty$ in tortoise coordinates
\begin{equation}
    \delta V_{\rm eff}(r)\to \delta V_{\rm eff}(r) \frac{1}{2}\left[1 + \tanh\left( \frac{r_* - r_{\rm cut}}{\Delta} \right) \right].
\end{equation}
This treatment physically corresponds to the redshift freezing mechanism. 
As the redshift factor $f(r) \to 0$, high-frequency fluctuations deep within the horizon are effectively screened from the dynamical response window for a distant observer.

\section{Numerical simulation and scaling analysis}

In our numerical implementation, we fix the benchmark intensity at $\epsilon=10^{-3}$. 
This choice is physically motivated by the need for a high-fidelity geometric benchmark. 
By simulating a regime where the effective potential perturbations are macroscopic yet distinct from the background, we ensure that the scattering signals are sufficiently robust against numerical noise to resolve the precise dynamical response to the horizon structure.
This benchmark approach allows us to isolate the geometric effects of the correlation length $L_c$ from the fluctuation intensity. 
Crucially, the dynamical features extracted at this benchmark level can then be analytically extrapolated across the full physical range of $\epsilon$—scaling down to the quantum foam regime or up to classical-level deformations—governed by the perturbative linearity of the scattering amplitude.

With the intensity fixed, we systematically scan the correlation length $L_c$ from microscopic scales ($10^{-3}M$) to macroscopic scales ($10^2 M$). 
This scan covers distinct physical scenarios: the limit $L_c \ll M$ represents incoherent stochastic fluctuations (quantum foam), while $L_c \sim M$ corresponds to coherent geometric deformations induced by long-range structures.

\begin{figure}[!htb]
    \centering
    \subfigure[The ringdown waveforms $|\Psi(t)|$ for correlation lengths $L_c/M \in \{0.01, 0.1, 1, 10, 100\}$. The almost perfect overlap demonstrates the macroscopic stability of the causal structure against microscopic deformation.]{\includegraphics[width=0.9\linewidth]{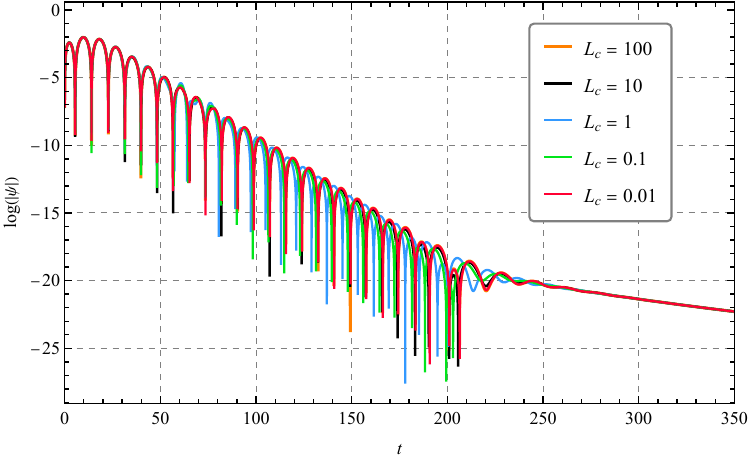}\label{fig:waveform-sub}}
    \subfigure[The residuals $\log |\Psi - \Psi_{\rm GR}|$ highlight two key phenomena: the amplitude follows an inverted U-shaped trend with a peak at the resonance scale $L_c \sim M$ (blue line), and the residual oscillation frequency matches the primary waveform frequency, indicating re-scattering origins.]{\includegraphics[width=0.9\linewidth]{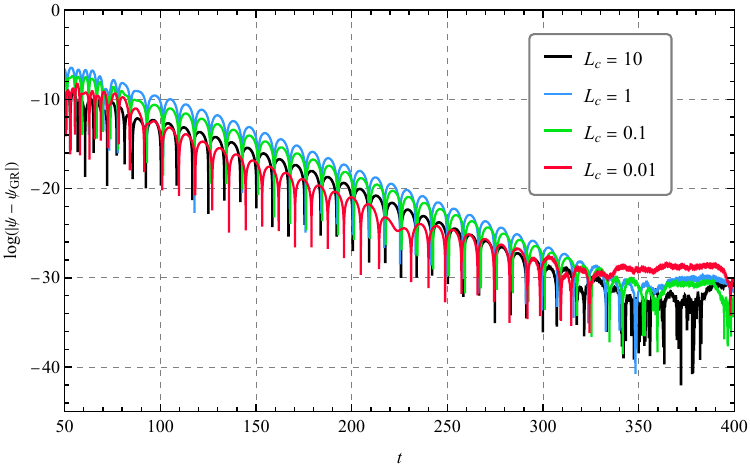}\label{fig:residue}}
    \caption{Time-domain evolution and residual analysis under fluctuations.}
    \label{fig:waveform}
\end{figure}

We employ a second-order accurate characteristic finite-difference evolution on the null grid $u=t-r_*$ and $v=t+r_*$ to solve the perturbed wave equation \cite{Gundlach:1993tp}.
To capture the fine structures of the potential barrier at the microscopic scale $L_c$, we utilize an extremely fine tortoise coordinate grid $\Delta r_* = 10^{-3} M$. 
The initial data are specified by a Gaussian packet on the ingoing null axis,
\begin{equation}
\Psi(u,0)=0,\qquad \Psi(0,v)=A_0 \exp\left[-\frac{(v-v_c)^2}{2\sigma^2}\right],
\end{equation}
with $v_c = 10 M$ and $\sigma = M$, and the waveform is extracted at $r_* = 80 M$
\footnote{
In practice, when analyzing the time-domain waveforms, the minimum value of $L_c$ we actually adopt is $10^{-2} M$ (instead of reaching $10^{-3} M$ mentioned above). 
According to the Nyquist sampling theorem, the step size of the finite-difference time-domain $\Delta r_*$ must be smaller than $L_c/2$ to capture all details of the oscillatory potential. 
Since $\Delta r_* = 10^{-3} M$ is used in our simulations, further reducing $\Delta r_*$ to match smaller $L_c$ would result in an extremely long simulation time.}
.
The numerical results reveal a non-intuitive feature of waveform stability. 
In the microscopic correlation case $L_c = 0.01 M$, although the spatial derivative of the effective potential $\delta V$ leads to a violent deviation of the local potential barrier, the generated GW ringdown signals match the classical Schwarzschild background almost perfectly in both primary frequency and damping rate.
This phenomenon is illustrated in Fig.\ \ref{fig:waveform}. Fig.\ \ref{fig:waveform-sub} displays the ringdown waveforms $\Psi(t)$ for various values of $L_c$. 
Here we adopt the case of $L_c=100M$ as the reference branch. 
This is mathematically justified by the structure of Eq.\ \eqref{eq:delta-V}, since the perturbation is dominated by the gradient term $\dif \eta/\dif r$, at macroscopic scales ($L_c \gg M$), the fluctuation scales as $\epsilon/L_c \to 0$, rendering the potential indistinguishable from the classical Schwarzschild barrier.
It can be observed that even when the potential barrier is violently shattered, the primary structures of the waveforms remain highly overlapped.
This macroscopic robustness suggests that GW probes, acting as long-wavelength detectors, possess an inherent shielding ability against geometric details below their wavelength scale.

To quantitatively explore the impact of fluctuations at different scales, Fig.\ \ref{fig:residue} further provides the residual analysis $\log |\Psi - \Psi_{\rm GR}|$. 
Rigorous convergence tests confirm that the numerical discretization error, which scales as $O(\Delta r_*^2) \sim O(10^{-6})$ for the second-order characteristic finite-difference scheme, is several orders of magnitude smaller than the signal, supporting the physical authenticity of these residuals.
We identify two core physical characteristics. 
First, the residual amplitude exhibits a distinct non-monotonic evolution with $L_c$: the residual reaches its maximum at $L_c = M$, while the deviation attenuates as the scale either decreases ($L_c = 0.1 M,\ 0.01 M$) or increases ($L_c = 10 M$). 
This confirms the existence of a resonance window where the correlation scale strongly couples with the gravitational wavelength of the dominant mode that governs the GW ringdown waveform ($\lambda_{\rm GW} \sim M$), and quantifies the scale at which microscopic structures achieve observational indistinguish ability.
Second, the residual signals manifest an oscillatory frequency highly consistent with the primary wave, confirming that the residuals are not unstructured background noise but originate from the re-scattering of the primary pulse by the stochastic potential.

In summary, the time-domain simulations demonstrate a remarkable macroscopic robustness: the global signal remains largely immune to the violent shattering of the local potential, yielding the notable smoothness of Fig.\ \ref{fig:residue} dictated by spatially integrated phase information. 
However, this phenomenological stability poses a theoretical puzzle: given the mathematical sensitivity of quasinormal modes to potential perturbations, what physical mechanism shields the ringdown waveform from this high-frequency spatial noise? To resolve this quantitatively and identify the microscopic origin of the stability, we must shift from the time-domain evolution to a frequency-domain scattering analysis.

\section{Frequency domain analysis and phase averaging mechanism}

\begin{figure}
    \centering
    \subfigure[Reflectivity $|\mathcal{R}|$ exhibits extreme inertia against microscopic potential shattering. ]{\includegraphics[width=0.9\linewidth]{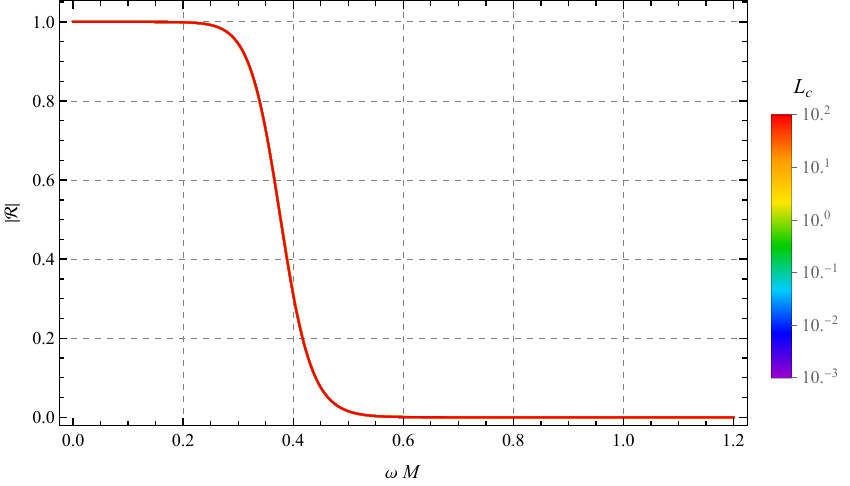}\label{fig:reflectivity}}
    \subfigure[Scattering phase shift $\delta(\omega)$ converges in the low-frequency regime ($\omega M \ll 1$), demonstrating the inherent suppression of short-wavelength modes by the wave operator.]    {\includegraphics[width=0.9\linewidth]{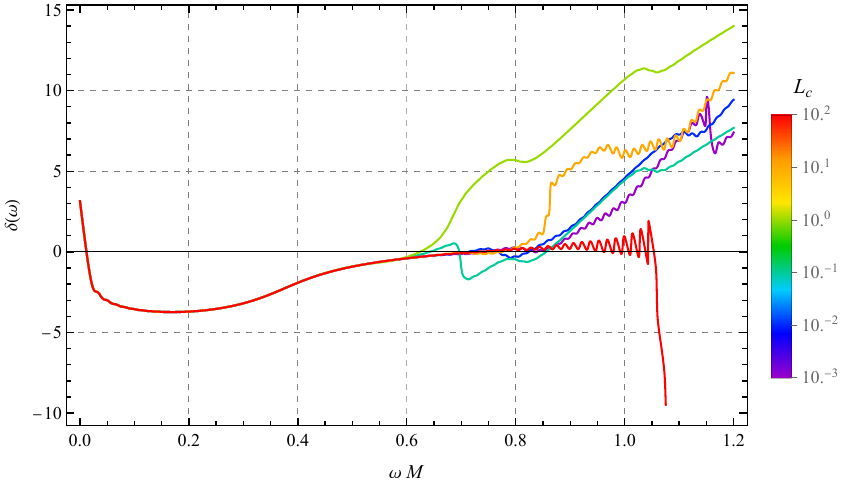}\label{fig:phase-shift}}
    \caption{Frequency-domain scattering response for a Schwarzschild BH with $\epsilon = 10^{-3}$.}
    \label{fig:freq-domain}
\end{figure}

To gain a deeper understanding of the physical origins of waveform robustness, we shift our focus from time-domain evolution to frequency-domain scattering characteristics. 
Although the QNM spectrum is highly sensitive to perturbations of the potential, the distribution of the total spectral weight typically exhibits greater resilience. 
The scattering phase shift $\delta(\omega)$, as an observable reflecting the cumulative effect of background geometry on wave phases, directly characterizes the impact of structural uncertainties near the horizon on GW propagation. 
We consider the perturbed Regge-Wheeler equation and rewrite it in a non-homogeneous form
\begin{equation}
    \left[ \frac{\dif^2}{\dif r_*^2} + \omega^2 - V_{\rm RW}(r_*) \right] \Psi(\omega, r_*) = \delta V_{\rm eff}(r_*)\Psi(\omega, r_*),
\end{equation}
where $V_{\rm RW}$ is the classical Regge-Wheeler potential. 
Utilizing the background Green's function $G(r_*, r_*^{\prime})$, which satisfies the boundary conditions for the unperturbed operator, we can express the scattered wave solution via the first-order Born approximation \cite{Oshita:2024wgt}
\begin{equation}
    \Psi(r_*) = \Psi_0(r_*) + \int_{-\infty}^{+\infty}\dif r_*^{\prime}\  G(r_*, r_*^{\prime}) \delta V_{\rm eff}(r_*^{\prime}) \Psi_0(r_*^{\prime}),
\end{equation}
where $\Psi_0$ is the incident wave solution in the GR background.
To derive the response of the scattering coefficients to the perturbation, we introduce the solution to the unperturbed equation, $\Psi_0$, and construct the Wronskian identity $\mathcal{W}[\Psi, \Psi_0] = \Psi \Psi_0^{\prime} - \Psi^{\prime} \Psi_0$. 
Utilizing integration by parts, we establish the following integral relation
\begin{equation}\label{eq:Wronskian}
    \mathcal{W}[\Psi, \Psi_0] \big|_{-\infty}^{+\infty} = \int_{-\infty}^{+\infty}\dif r_*\ \delta V_{\rm eff}(r_*) \Psi(\omega, r_*) \Psi_0(\omega, r_*).
\end{equation}
In the scattering problem, we define the following boundary conditions: 
at the horizon ($r_* \to -\infty$), the wave satisfies the purely outgoing condition $\Psi \sim \mathcal{T} \me^{-\mi\omega r_*}$;
at infinity ($r_* \to +\infty$), the wave manifests as a superposition of a unit-amplitude incident wave and a reflected wave $\Psi \sim \me^{-\mi\omega r_*} + \mathcal{R} \me^{\mi\omega r_*}$, where $\mathcal{R}$ and $\mathcal{T}$ denote the reflection and transmission coefficients, respectively. 
Evaluating the boundary terms on the left side of Eq.\ \eqref{eq:Wronskian}, the Wronskian contribution at the horizon vanishes due to the proportionality of the solutions, while the contribution at infinity is proportional to the deviation of the reflection coefficient. 
Under the first-order Born approximation, we parametrize the reflectivity as $\mathcal{R} = |\mathcal{R}| \me^{2 \mi \delta}$. 
Assuming that the amplitude change induced by $\delta V$ is negligible compared to the phase shift (also shown in Fig.\ \ref{fig:reflectivity}), the variation in the reflection coefficient can be linearized as $\mathcal{R} - \mathcal{R}_0 \simeq 2\mi \mathcal{R}_0 \Delta \delta$. 
Substituting the boundary values yields
\begin{equation}
    2 \mi \omega(\mathcal{R}_0 - \mathcal{R}) \simeq \int_{-\infty}^{+\infty} \dif r_*\ \delta V_{\rm eff}(r_*) \Psi_0^2(\omega, r_*).
\end{equation}
Consequently, the phase deviation $\Delta \delta$ produced by the microscopic fluctuation $\delta V_{\rm eff}$ is explicitly expressed as a spatial integral over the stochastic potential field
\begin{equation}\label{eq:phase-shift}
    \Delta \delta(\omega) \simeq \frac{1}{4\omega \mathcal{R}_0} \int_{-\infty}^{+\infty}\dif r_*\ \delta V_{\rm eff}(r_*) \Psi_0^2(\omega, r_*). 
\end{equation}
Using Eq.\ \eqref{eq:delta-V}, the stochastic correction can be reorganized schematically as $\delta V_{\rm eff}(r_*)=\epsilon[U_0(r_*)\zeta(r_*)+U_1(r_*)\partial_{r_*}\zeta(r_*)]$, where $U_0$ and $U_1$ are smooth background functions. 
Introducing the response weight
\begin{equation}
K_\omega(r_*) \equiv \frac{\Psi_0^2(\omega,r_*)}{4\omega \mathcal{R}_0},
\end{equation}
and integrating the derivative term by parts, Eq.\ \eqref{eq:phase-shift} can be rewritten as
\begin{equation}
\Delta\delta(\omega) \simeq \epsilon \int_{-\infty}^{+\infty} \dif r_*\ J_\omega(r_*)\zeta(r_*),
\end{equation}
where
\begin{equation}
J_\omega(r_*) \equiv U_0(r_*)K_\omega(r_*)+\partial_{r_*}[U_1(r_*)K_\omega(r_*)].
\end{equation}
For a zero-mean stochastic field, the leading ensemble-averaged shift vanishes, while the variance is controlled by the overlap between the stochastic power spectrum $P_{L_c}$ of $\zeta(r_*)$  and the response kernel,
\begin{equation}
\mathrm{Var}[\Delta\delta(\omega)] = \epsilon^2 \int \frac{\dif k}{2\pp}\ P_{L_c}(k) |\widetilde{J}_\omega(k)|^2,
\end{equation}
where $\widetilde{J}_\omega(k) = \int \dif r_* \ \me^{-\mi k r_*} J_\omega(r_*)$.
This makes the phase-averaging mechanism quantitative: the observed signal is governed by a coarse-grained overlap on the real axis rather than by the pointwise jaggedness of $\delta V_{\rm eff}$. 
In particular, the smooth residuals in Fig.\ \ref{fig:residue} arise because the detector probes this kernel-averaged response, not the ultraviolet extrema of the barrier.
This phase-integral perspective provides a physical resolution to the waveform instability. 
The observable ringdown signal is determined by the scattering phase shift on the real axis $\omega\in\mathbb{R}$ \cite{Kyutoku:2022gbr}. 
Eq.\ \eqref{eq:phase-shift} reveals that the Green's function integral suppresses high-wavenumber contributions: the microscopic phase deviations induced by the rapidly oscillating potential $\delta V_{\rm eff}$ undergo efficient destructive interference, thereby shielding the macroscopic waveform from ultraviolet instability.

Eq.\ \eqref{eq:phase-shift} reveals the mathematical essence of the phase averaging: 
For the fundamental GWs that dominate the ringdown signal, the characteristic wavelength satisfies $\lambda_{\rm GW} \sim M$. 
Under the scale $\lambda_{\rm GW} \gg L_c$, the wavefront traverses a vast number of stochastic fluctuation units as it propagates through the random potential field. 
Since $\delta V_{\rm eff}$ oscillates rapidly at the scale of the correlation length $L_c$ and its ensemble average $\langle \delta V_{\rm eff} \rangle$ vanishes, the integral term in Eq.\ \eqref{eq:phase-shift} is suppressed by the high-frequency oscillating phases according to the Riemann-Lebesgue lemma. 
This implies that microscopic phase deviations produced by different local extrema undergo efficient destructive interference along the integration path.
The inverse operator of the wave equation (the Green's function integral) essentially averages out spatial fluctuations, effectively filtering out the contribution of ultraviolet perturbations to the macroscopic phase accumulation.

The robustness of this frequency response is quantitatively demonstrated in Fig.\ \ref{fig:freq-domain}.
Fig.\ \ref{fig:reflectivity} presents the scan of the reflectivity $|\mathcal{R}|$ as a function of frequency $\omega$, revealing its extreme inertia toward microscopic disorder. 
Even when the potential barrier is severely shattered, the overall ratio of energy transmission and reflection remains close to the classical GR prediction. 
Fig.\ \ref{fig:phase-shift} illustrates the evolution of the scattering phase shift $\delta(\omega)$. 
In the low-frequency (long-wavelength) regime, the phase shift curves for different correlation lengths $L_c$ collapse onto a universal profile. 
The stochastic signatures of the horizon structure become distinguishable only in the high-frequency (short-wavelength) regime.
This phenomenon demonstrates that the dynamical response of the BH exerts a significant smoothing effect on microscopic geometric details.

\begin{figure}[!htb]
    \centering
    \includegraphics[width=\linewidth]{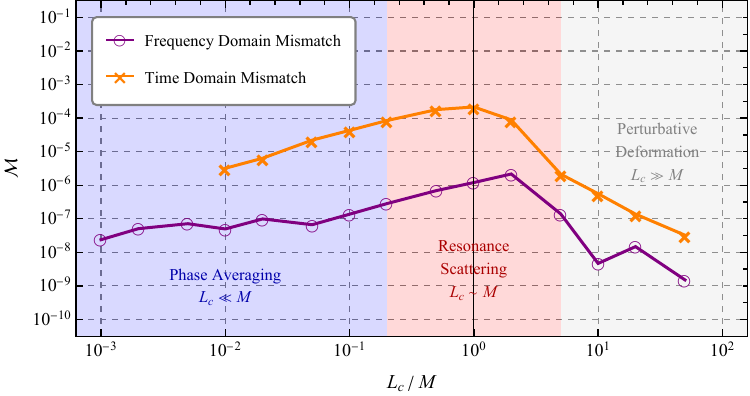}
    \caption{Scaling of the mismatch $\mathcal{M}$ as a function of correlation length $L_c$. 
    The ``inverted U-shaped" profile characterizes the resonance window near $L_c \sim M$ and the linear suppression of observational deviations in the microscopic limit ($L_c \ll M$), quantifying the resolution limit of the horizon.}
    \label{fig:mismatch}
\end{figure}

To quantify the dependence of observational deviations on the characteristic scale of the horizon structure, we present the scaling behavior of the mismatch $\mathcal{M}$ as a function of the correlation length $L_c$ in Fig.\ \ref{fig:mismatch}.
Here, the mismatch $\mathcal{M}$ is employed to quantitatively characterize the indistinguishability between the perturbed waveform and the classical GR reference. 
Considering the response characteristics of gravitational-wave detectors across different physical dimensions, we define the time-domain mismatch $\mathcal{M}_t$ and the frequency-domain mismatch $\mathcal{M}_f$ respectively
\begin{subequations}
    \begin{equation}
        \mathcal{M}_t = 1 - \frac{|\langle \Psi, \Psi_{\rm GR} \rangle_t|}{\sqrt{\langle \Psi, \Psi \rangle_t \langle \Psi_{\rm GR}, \Psi_{\rm GR} \rangle_t}},
    \end{equation}
    \begin{equation}
        \mathcal{M}_f = 1 - \frac{|\langle \tilde{\Psi}, \tilde{\Psi}_{\rm GR} \rangle_\omega|}{\sqrt{\langle \tilde{\Psi}, \tilde{\Psi} \rangle_\omega \langle \tilde{\Psi}_{\rm GR}, \tilde{\Psi}_{\rm GR} \rangle_\omega}}.
    \end{equation}
\end{subequations}
In these expressions, $\displaystyle \langle A, B \rangle_t = \int \dif t\ A(t) B^*(t)$ denotes the overlap integral in the time domain. 
For the frequency-domain mismatch $\mathcal{M}_f$, the inner product $\displaystyle \langle \tilde{A}, \tilde{B} \rangle_\omega = \int_{0}^{\omega_{\rm cut}} \dif \omega\ \tilde{A}(\omega) \tilde{B}^*(\omega)$ is specifically integrated over the infrared part, with a cutoff frequency of $\omega_{\rm cut} = 0.6/M$
\footnote{
The results are insensitive to the specific choice of $\omega_{\rm cut}$ as long as the integration region excludes the ultraviolet regime ($\omega_{\rm cut} \lesssim 0.6/M$).
}
. 
The choice of this frequency band is intended to focus on the coherent scattering contributions in the low-frequency regime, where modifications to the background geometry are most pronounced, while avoiding interference from high-frequency incoherent noise. 
A mismatch value closer to zero indicates a diminished impact of microscopic fluctuations on macroscopic observations.

\begin{figure}[!htb]
    \centering
    \includegraphics[width=0.9\linewidth]{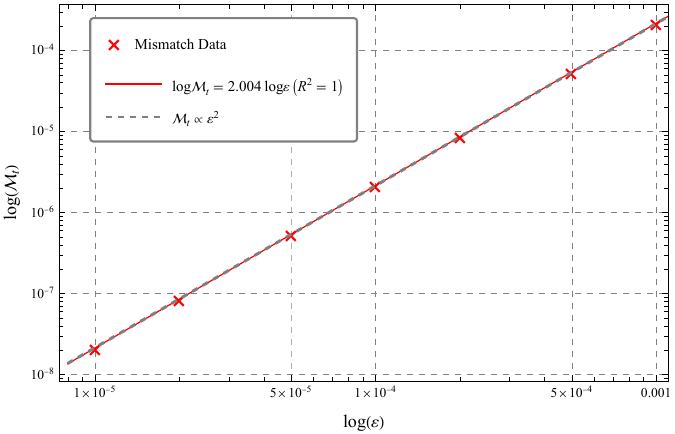}
    \caption{Validation of the perturbative scaling law. 
    The waveform mismatch $\mathcal{M}_t$ is plotted against the fluctuation intensity $\epsilon$ on a logarithmic scale. 
    The red crosses represent numerical results obtained from our time-domain simulations at the resonance scale $L_c \sim M$.
    The dashed gray line indicates the theoretical prediction $\mathcal{M} \propto \epsilon^2$.
    The excellent agreement (slope $\simeq 2.0$) across two orders of magnitude confirms that the scattering process remains in the linear response regime even up to $\epsilon \sim 10^{-3}$, justifying our extrapolation strategy.}
    \label{fig:scaling-law}
\end{figure}

The mismatch profile exhibits a characteristic 'inverted U-shaped' resonance: the deviation is maximized when $L_c \sim M$, a scale at which the horizon structure becomes commensurable with the gravitational wave frequency; conversely, when $L_c \ll M$, the mismatch is suppressed as $L_c$ decreases.
With the dependence on correlation length $L_c$ elucidated by the resonance profile, we now address the fluctuation intensity $\epsilon$ to map the entire parameter space. 
We employ a hybrid strategy based on analytical extrapolation: while $L_c$ determines the geometric shape of the deviation curve, the overall observability is linearly governed by the intensity squared. 
As verified in Fig.\ \ref{fig:scaling-law}, our numerical tests confirm that the mismatch amplitude follows the perturbative scaling law $\mathcal{M}\propto\epsilon^{2}$ strictly.
This linearity implies that the ``inverted U-shaped" curve in Fig.\ \ref{fig:mismatch} represents a universal geometric profile; varying $\epsilon$ simply shifts the entire curve vertically without altering its resonant shape.

We can now quantitatively map these results to realistic observational scenarios. 
Considering a typical detection threshold of $\mathcal{M}_{\rm det} \sim 10^{-5}$ for next-generation detectors, our scaling relation $\mathcal{M} \simeq 200 \epsilon^2$—derived from the resonance peak at $L_c \sim M$—imposes a stringent detectability lower bound $\epsilon_{\rm det} \gtrsim 2 \times 10^{-4}$.
This magnitude imposes a stringent physical bound, implying that for any horizon-scale structure to be observable via ringdown spectroscopy, it must satisfy two constraints:
1) Macroscopic coherence ($L_c \sim M$), meaning the structure must possess long-range correlations to trigger geometric resonance; and 
2) Macroscopic intensity ($\epsilon \sim 10^{-4}$), requiring the effective geometric deformation to be significantly larger than the intrinsic Planck scale.

This result clarifies the domain of validity for our stochastic framework and its connection to quantum gravity candidates. 
It is crucial to distinguish the macroscopic fluctuations modeled here from standard microscopic quantum foam. 
While the original Einstein-Langevin equation is formally derived for Planck-scale quantum fluctuations ($\epsilon \sim l_{\rm P}/M \sim 10^{-30}$), the observability threshold derived above ($\epsilon \sim 10^{-4}$) corresponds to macroscopic geometric deformations that exceed the Planck scale by thirty orders of magnitude. 
In this regime, the stochastic potential $\delta V_{\rm eff}(r)$ should not be interpreted as quantum foam, but rather as an effective field theory parameterization of the spatial complexity inherent in the horizon structure.

This framework finds natural application in the context of the fuzzball proposal, where individual microstate geometries are expected to break the classical symmetries of the standard GR geometry, leading to a rich multipolar structure \cite{Bianchi:2020bxa}.
Although the exact configuration varies across the immense phase space of $\me^{S_{\rm BH}}$ microstates, the effective potential experienced by a ringdown wave scattering off a generic microstate can be phenomenologically modeled as a random field with a characteristic correlation length.
In this picture, $\epsilon$ quantifies the root-mean-square amplitude of the geometric deviation from the ensemble average (the classical GR background), while $L_c$ characterizes the coherence scale of these deformations.
Consequently, our analysis implies a statistical shielding mechanism: the vast majority of the microstate ensemble—characterized by high-entropy, incoherent structure—remains observationally indistinguishable from classical BHs due to phase cancellation. 
A detectable signal would strictly require the horizon geometry to exhibit macroscopically coherent variance ($L_c \sim M$) with classical-level intensity ($\epsilon \gtrsim 10^{-4}$), a signature corresponding to coherent states or exotic compact objects with significant surface anisotropy.

\begin{figure}
    \centering
    \subfigure[Reflectivity $|\mathcal{R}|$ as a function of frequency. The scattering profile remains remarkably stable across broad scales, with the shift at $L_c = 0.001M$ marking the validity bound of the stepwise approximation in mimicking a continuous background]{\includegraphics[width=0.9\linewidth]{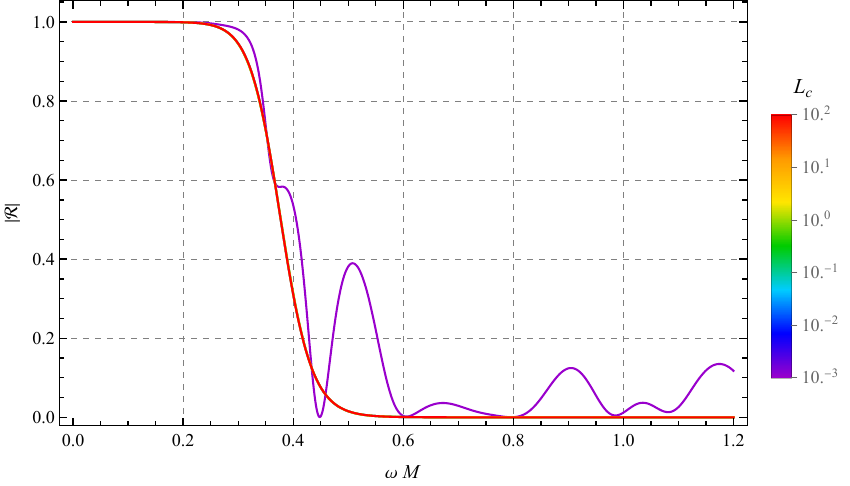}}
    \subfigure[Scattering phase shift $\delta(\omega)$ for various correlation lengths $L_c$. High convergence is maintained for $L_c \geq 0.005M$, while the divergence at $L_c = 0.001M$ (purple line) characterizes the limit where systematic discretization bias—arising from extremely high-frequency sampling of the potential minima.]{\includegraphics[width=0.9\linewidth]{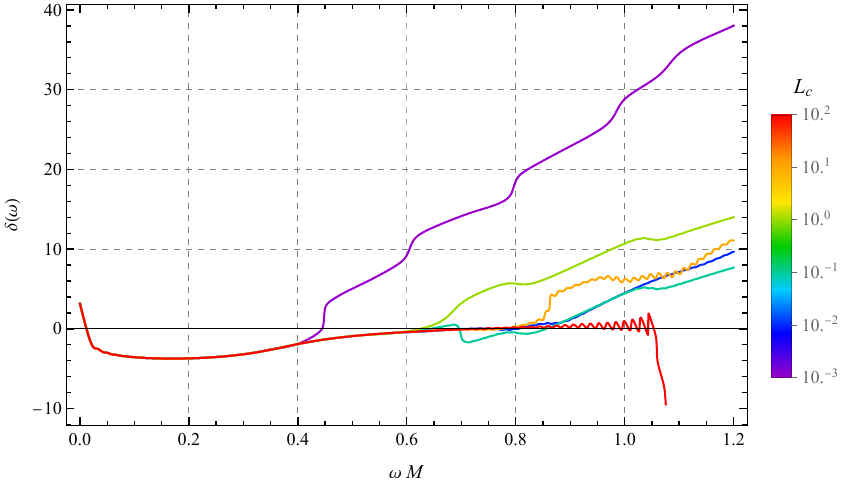}}
    \caption{Frequency-domain response under the conservative inscribed step approximation scheme of the potential with $\epsilon = 10^{-3}$.}
    \label{fig:freq-domain-step}
\end{figure}

Furthermore, since the convolution in Eq.\ \eqref{eq:convolve} produces a continuous colored-noise realization, it is useful to verify explicitly that the stability mechanism does not rely on local smoothness. 
We therefore introduce the conservative inscribed step approximation of the potential as a complementary worst-case test, in close analogy with recent piecewise-step analyses of waveform stability \cite{Wu:2025sbq}.
This scheme simulates the response in the highly non-smooth limit by discretizing the stochastic potential into steps of width $L_c$ and extracting local minima.
This construction creates a worst-case scenario by introducing derivative discontinuities at step edges and imposing a systematic suppression on the potential barrier, thereby rigorously testing whether the macroscopic stability relies on the local differentiability of the underlying horizon geometry.
As shown in Fig.\ \ref{fig:freq-domain-step}, across a broad range of $L_c \ge 0.005M$, the reflectivity and phase shift maintain high consistency with Fig.\ \ref{fig:freq-domain}. 
This indicates that even under discretized, non-smooth perturbations, the integrative effect of long-wavelength probes on microscopic fluctuations dominates the observational response.

Notably, as the correlation length further decreases to $L_c = 0.001M$, the inscribed step approximation exhibits identifiable frequency-domain deviations.
We verified that this deviation is numerically convergent, confirming it as a discretization artifact inherent to the stepwise approximation of the stochastic field.
This deviation identifies a validity bound specific to the stepwise approximation in the microscopic limit. 
Unlike larger correlation lengths ($L_c \ge 0.01M$) where the step potential effectively mimics a continuous profile, at $L_c \sim 10^{-3}M$, the high density of artificial step discontinuities introduces spurious coherent scattering. 
This systematic bias arises because the inscribed step model effectively replaces the smooth random field with a jagged barrier, generating non-physical high-wavenumber noise that dominates only in this deep ultraviolet regime.
Consequently, the divergence observed here reflects the breakdown of the discrete step model itself, rather than a failure of the physical phase averaging mechanism.

\section{Conclusion and discussion}

In this work, we have systematically examined the robustness of BH ringdown to stochastic horizon-scale structure, utilizing an effective field framework to parametrize the geometric fluctuations.
Although pseudospectrum analysis based on non-Hermitian operators suggests that the QNM spectrum of BHs is extremely sensitive to potential perturbations \cite{Jaramillo:2020tuu}, our time-domain waveform and phase-shift analysis proves that this sensitivity is significantly attenuated in the scattering response on the real frequency axis, as the wave packet perceives the spatial integration of the potential. 
This supports the hypothesis of greybody factor stability \cite{Ianniccari:2024ysv}: while individual QNMs are mathematically sensitive to perturbations, their collective superposition can still leave the observable signal stable \cite{Oshita:2023cjz}. 
We advance this understanding by identifying the phase averaging as the physical origin of this robustness.
We demonstrate that the stochastic horizon acts as a diffusive scattering medium; microscopic rugosity are effectively smoothed out by the wave propagation integral, rendering the ringdown signal insensitive to short-wavelength fluctuations below the resolution limit determined by the probing wavelength.

Building on this mechanism, we establish the constraint on the stochastic structure of BH for observability.
The resonance characteristics identified in our analysis impose strict constraints on any detectable signal:
for a deviation to exceed the noise threshold of next-generation detectors ($\mathcal{M} \gtrsim 10^{-5}$), the horizon structure must possess both macroscopic spatial coherence ($L_c \sim M$) and classical-level intensity ($\epsilon \gtrsim 10^{-4}$). 
This criterion quantitatively rules out the observability of incoherent, high-entropy quantum foam or standard vacuum fluctuations via ringdown spectroscopy, as their lack of spatial correlation leads to drastic suppression via phase cancellation. 
Consequently, any observed non-GR ringdown signal in this channel would point toward macroscopically coherent horizon structures that sustain classical geometric deformations.
The present work is intentionally a proof-of-principle analysis in the Schwarzschild background.
Nevertheless, even for realistic spinning BHs, the underlying phase-averaging argument is controlled by the ratio between the structural scale and the probing wavelength, so away from extremality one expects the peak sensitivity to track the dominant Kerr ringdown wavelength,
\begin{equation}
L_c^{\rm peak}(a) \sim \lambda_{220}(a) \simeq \frac{2\pp}{\Re \omega_{220}(a)},
\end{equation}
where $a$ denotes the dimensionless spin parameter.
The short-correlation suppression persisting whenever $L_c \ll \lambda_{220}(a) \sim O(M)$. 
In this sense, generic spin should mainly shift the resonance window by an order-unity factor rather than remove the mechanism identified here.
Near extremality, however, higher-derivative corrections and effective field theory effects can be parametrically enhanced, and the corresponding Kerr ringdown shifts display special scaling behavior \cite{Horowitz:2023xyl,Boyce:2026rnn}.
A dedicated analysis is therefore required for the near-extremal regime.

These constraints clarify the interpretation of the horizon stochastic structure within our model. 
The shattering potential employed here serves as an effective field theory parameterization of the geometric variance inherent in the BH microstate ensemble. 
Our results suggest that typical thermal states of the ensemble, characterized by microscopic quenched disorder, remain observationally indistinguishable from the classical GR situation due to the phase cancellation of short-wavelength fluctuations. 
Conversely, significant ringdown deviations can only arise from macroscopically coherent states, such as fuzzball geometries or exotic compact objects characterized by long-range spatial correlations and surface anisotropy. 
In this scenario, the randomness of the potential captures the spatial complexity arising from the non-trivial topology of the underlying coherent structure, providing a phenomenological bridge between fundamental quantum gravity descriptions and macroscopic observables.
The robustness of this phenomenological picture is further reinforced when considering the temporal evolution of the microstates.
Any rapid temporal fluctuations would dynamically average out over the wave's interaction timescale, further suppressing the observable response.
The present quenched treatment therefore serves as a robust conservative upper bound for the impact of zero-mean stochastic structures, leaving the full explicitly time-dependent scattering as an interesting extension.

Furthermore, the framework proposed in this work offers a new perspective on the search for GW echoes. 
The geometric structures induced by the stochastic potential can be viewed as diffuse scattering centers. 
Unlike the specular reflection expected from a perfectly reflective surface, the random phase delays generated by these fluctuations lead to severe decoherence of echo signals during multiple round trips in the cavity. 
This suggests that for rough or structured horizons, the coherent echo trains may be smeared into indistinguishable broadband noise, potentially explaining the null results in current echo searches within LIGO/Virgo data. 
This highlights the necessity of incorporating phase coherence loss into future waveform templates for exotic compact objects.

Looking ahead, a natural extension of this work is to incorporate rotation, where the interplay between the correlation scale of the horizon geometry and the ergoregion instability may introduce novel dynamical features. 
This is especially meaningful for high-spin and near-extremal Kerr backgrounds.
Additionally, mapping specific coherent microstate geometries directly to the statistical parameters $(\epsilon, L_c)$ of our effective model would facilitate more precise constraints on quantum gravity theories using future high-precision GW data.

\vspace{5pt}
\noindent
\section*{Acknowledgments}
This work is supported by the National Natural Science Foundation of China No.~12475067 and No.~12235019.

\appendix

\bibliographystyle{apsrev4-1}
\bibliography{main}

\end{document}